\documentclass{emulateapj}
\shorttitle{MHD Poynting flux in solar photospheric magneto-convection simulations}
\shortauthors{S.~Shelyag, M.~Mathioudakis, F.P.~Keenan}

\begin{document}
\title{Mechanisms for MHD Poynting flux generation in simulations of solar photospheric magneto-convection}

\author{S.~Shelyag, M.~Mathioudakis, F.P.~Keenan}

\affil{Astrophysics Research Centre, School of Mathematics and Physics, Queen's University, Belfast,
BT7 1NN, Northern Ireland, UK}

\date{01.01.01/01.01.01}

\begin{abstract}
We investigate the generation mechanisms of MHD Poynting flux in the magnetised solar photosphere.
Using radiative MHD modelling of the solar photosphere with initial magnetic configurations that differ in their
field strength and geometry, we show the presence of two different mechanisms for MHD Poynting flux generation 
in simulations of solar photospheric magneto-convection. The weaker mechanism is connected to vertical 
transport of weak horizontal magnetic fields in the convectively stable layers of the upper photosphere, while the 
stronger is the production of Poynting flux in strongly magnetised intergranular lanes experiencing 
horizontal vortex motions. These mechanisms may be responsible for the energy transport from the solar 
convection zone to the higher layers of the solar atmosphere.
\end{abstract}

\keywords{Sun: Photosphere --- Sun: Surface magnetism --- Plasmas --- Magnetohydrodynamics (MHD)}

\section{Introduction}
The magnetic coupling between the solar interior and the outer layers of the solar atmosphere plays a key role in the 
energy transport between these regions.  We have recently demonstrated \citep{shelyag2011} that vorticity generation 
in the intergranular lanes of magnetic solar plage region simulations is much more efficient compared to the non-magnetic 
case. These vortices have been detected in observations of the photosphere and chromosphere 
\citep[see e.g.][]{bonet1,wedemeyer1}. The magnetic vortex motions are highly significant as they can deposit large 
amounts of energy in the solar chromosphere \citep{moll1}, contribute to the production of Poynting flux and generate 
various types of MHD wave modes \citep{shelyagangeo,fedunangeo}. On the other hand, using a simulation setup that 
involved influx of weak horizontal magnetic field through the bottom boundary of the simulation domain, \citet{steiner1} 
have shown that the Poynting flux is generated through the advection of horizontal magnetic fields by vertical upflows in the 
upper, convectively stable layers of the solar photosphere. \citet{abbett1} have recently demonstrated the presence of 
Poynting flux in their radiative MHD simulations of an open-flux region of the solar atmosphere.

In this paper we use forward modelling techniques to examine the generation of Poynting flux in a model of the solar 
magneto-convection with different initial magnetic field strengths and geometries.  Our model includes the 
wavelength-dependent radiative transport and the partial ionisation of a plasma with solar chemical composition. 
We perform simulation runs with initial vertical magnetic fields of $10~\mathrm{G}$ and $200~\mathrm{G}$, as well as 
with changing polarity with unsigned flux of $200~\mathrm{G}$ and with weak horizontal 
magnetic field of $10~\mathrm{G}$, advected into the numerical domain by convective upflows. The 
effectiveness of Poynting flux generation is analysed in these simulations and we find that Poynting flux generation by vortices, 
outlined by \citet{shelyagangeo}, is more pronounced irrespective of the magnetic configuration. We use a mixed-polarity 
region to study the dependence of Poynting flux generation relative to the degradation of activity.

\section{Simulation setup}

\begin{figure*}
\epsscale{1.0}
\plotone{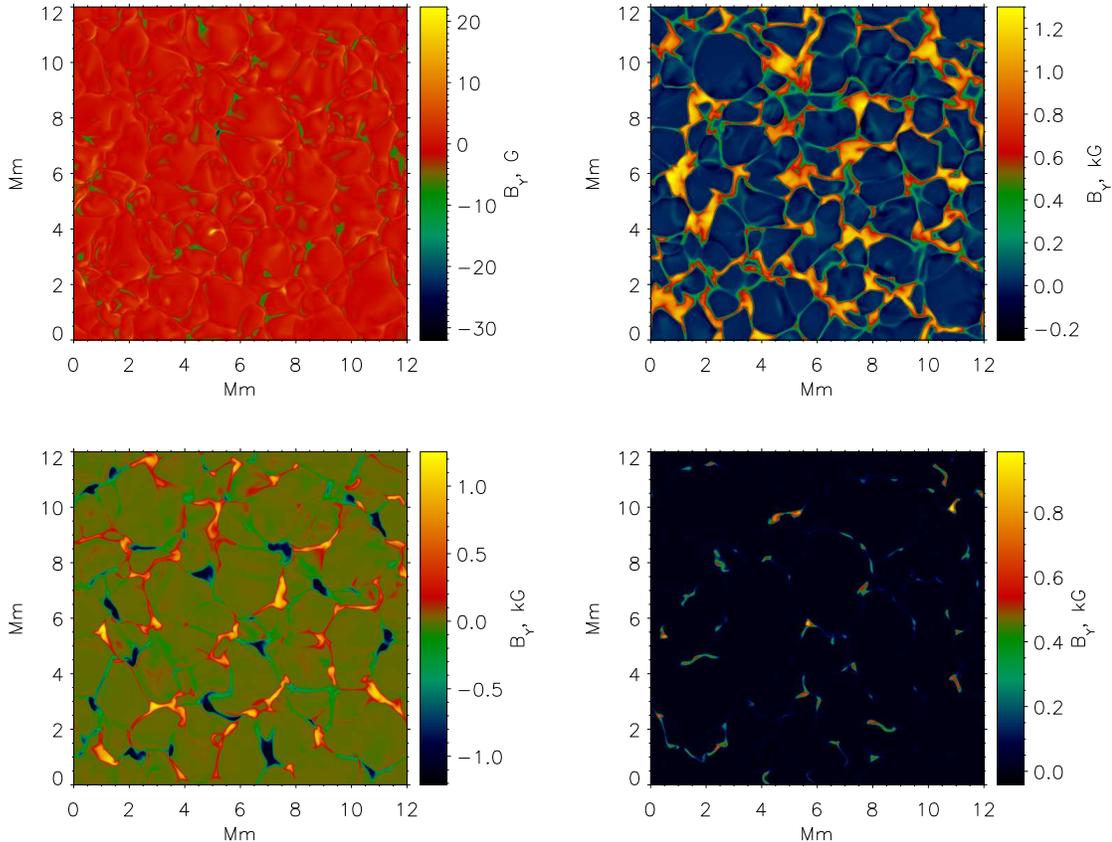}
\caption{Vertical component of the magnetic field measured at the photospheric level in four simulations that differ in 
their initial field strength and geometry. Top left: horizontal magnetic field of $10~\mathrm{G}$ advected through the 
bottom boundary of the numerical domain; top right: $200~\mathrm{G}$ initial vertical magnetic field; bottom left:
mixed-polarity initial vertical $200~\mathrm{G}$ magnetic field, bottom right: $10~\mathrm{G}$ initial vertical magnetic 
field simulation. The images have been taken approximately $800~\mathrm{s}$ after the magnetic field was introduced 
into the domain.}
\label{fig1}
\end{figure*}

Our numerical modelling of the solar magnetised photosphere is undertaken using the MURaM radiative MHD code 
\citep{voegler1}. The code is described in detail in \citet{voeglerphd} and it has been successfully used for a wide 
variety of applications in solar physics. Here we provide only a brief description of the code. It uses 4-th order 
central difference operators to calculate spatial derivatives, a 4-th order Runge-Kutta scheme to advance the solution 
in time, and hyperdiffusive operators to stabilise the solution against numerical instabilities. A non-grey radiative source
term is included in the MHD equations to account for wavelength-dependent radiative energy exchange in the simulated
photosphere, and a non-ideal equation of state is used, which takes into account partial ionisation of the 11 most abundant elements 
in the solar photosphere. Side boundary conditions are periodic, and the top boundary is closed for in- and 
outflows, while allowing horizontal motions of fluid and magnetic field lines. The bottom boundary condition is open for in- 
and outflows. In the simulations with an initial vertical magnetic field, magnetic field lines are forced to be vertical at the 
bottom boundary. We also use an alternative configuration of the bottom boundary condition, that allows the influx of 
horizontal magnetic field with a strength of  $10~\mathrm{G}$ in the convective upflow regions and leaves the outflows 
unaffected. This configuration has been described in more detail by  \citet{steinbc} and has been used in the simulations 
of \citet{steiner1}.    

We use a domain size of $12 \times 12~\mathrm{Mm^2}$ in the horizontal directions, and $1.4~\mathrm{Mm}$ in the vertical, 
resolved by $480 \times 480$ and 100 grid cells, respectively. The domain is positioned in such way that the 
continuum formation level is located approximately $800~\mathrm{km}$ above the bottom boundary. We use this setup 
to perform a series of simulations with $10~\mathrm{G}$ and $200~\mathrm{G}$ initial uniform vertical magnetic fields, 
as well as a simulation with initial vertical magnetic field of mixed (positive-negative) polarity with a net strength of 
$200~\mathrm{G}$, similar to that used by \citet{cameron1}. A simulation with a $10~\mathrm{G}$ horizontal magnetic 
field, advected with upflows into the domain though its bottom boundary, is also carried out.

Starting from a non-magnetic photospheric convection snapshot, we introduced these magnetic fields and boundary 
conditions in the domain and recorded sequences of magneto-convection snapshots over a few granular lifetimes.

\begin{figure*}
\epsscale{1.0}
\plotone{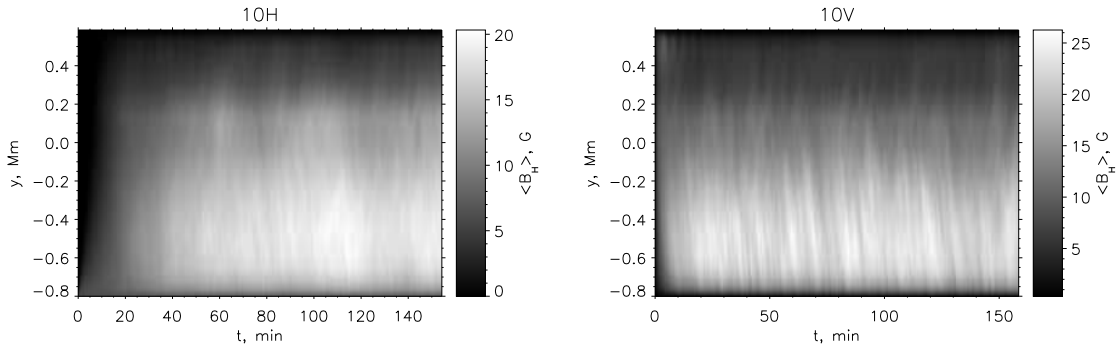}
\caption{Time evolution of average horizontal magnetic field strength as a function of height. Left panel: $10~\mathrm{G}$ 
horizontal magnetic field advected through the bottom boundary; right panel: a $10~\mathrm{G}$ initially vertical magnetic 
field simulation.}
\label{fig2}
\end{figure*}

In Fig.~\ref{fig1}, vertical magnetic field distribution $B_y \left(x, z\right)$ in the domain at the photospheric level is shown for four simulation 
runs of different field strength and geometry. The top left panel shows that the interaction of horizontal 
magnetic field influx with convective flows does not form strong intergranular magnetic flux concentrations, in contrast to the 
$10~\mathrm{G}$ and $200~\mathrm{G}$  initial vertical uniform magnetic field (top right and bottom right). Small-scale 
loop-like structures with magnetic fields of about $10-30~\mathrm{G}$ are formed instead due to the advection of magnetic 
field by granular upflows. The area covered by magnetic field concentrations in the $10~\mathrm{G}$ vertical case is significantly 
lower compared to that for $200~\mathrm{G}$, and the magnetic field strength in these concentrations is also somewhat lower. 
In the bottom left panel of Fig.~\ref{fig1} the distribution of the vertical component of magnetic field for the mixed-polarity simulation is shown. Here, 
strong magnetic flux concentrations with strengths of up to about $1.2~\mathrm{kG}$ are initially formed. However, the net 
magnetic flux decreases during the simulation as a result of opposite-polarity magnetic flux reconnection and cancellation 
\citep{cameron1}, and due to the magnetic diffusivity in this relatively low-resolution run. \citep[See e.g.][for a discussion of magnetic 
field generation and  decay speed dependence on the numerical grid resolution and effective magnetic Reynolds number]{voeglerdyn}.

\begin{figure}
\epsscale{1.0}
\plotone{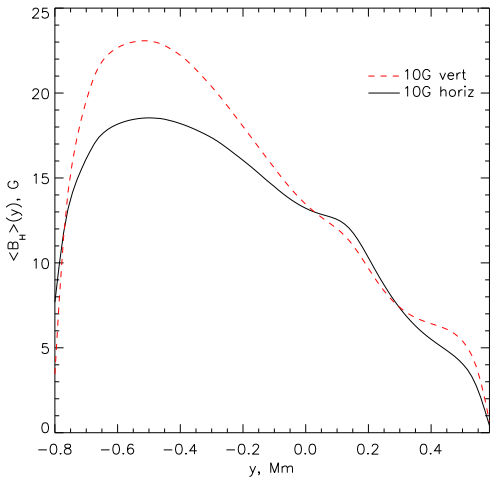}
\caption{Time average of the variation in the mean horizontal magnetic field with height. The dashed curve corresponds to the $10~\mathrm{G}$ 
initial vertical uniform magnetic field setup, while the black solid curve is the $10~\mathrm{G}$ horizontal magnetic field simulation.}
\label{fig3}
\end{figure}

The $10~\mathrm{G}$ initial vertical uniform magnetic field and $10~\mathrm{G}$ horizontal magnetic field simulations show significant
differences in the evolution of their average quantities. This is plotted in Fig.~\ref{fig2}, where the time evolution of the modulus of 
horizontal components of magnetic field is shown for the two cases. In the horizontal magnetic field simulation, it takes approximately 
15 minutes for the magnetic field to propagate from the bottom boundary of the domain to the photospheric level (marked by $y=0$ 
in the left plot of Fig.~\ref{fig2}). In agreement with the findings of \citet{steiner1}, the horizontal magnetic field is accumulated at a height 
of approximately $0.15~\mathrm{Mm}$ above the average continuum formation level $y=0$. This is visible as a small increase in the 
average horizontal magnetic field at this height. A similar plot for the $10~\mathrm{G}$ vertical field is shown in the right panel of 
Fig.~\ref{fig2}. In the latter case, the vertical magnetic field is redistributed by the convective flows within the first 5 minutes of the simulation, 
and the horizontal component of the magnetic field appears. There is also no clear layer where the horizontal magnetic field is 
accumulated. This is further demonstrated in Fig.~\ref{fig3}, where we plot the time averages of the dependence of the horizontal 
magnetic field as a function of height. The figure shows a local enhancement of the horizontal magnetic field strength at 
$y=0.15~\mathrm{Mm}$ for $10~\mathrm{G}$ horizontal field simulation, while for the vertical magnetic field simulation there is almost 
no enhancement.

\section{Poynting flux analysis}

We define the MHD Poynting flux vector as
\begin{equation}
\mathbf{F}=\frac{1}{4\pi} \mathbf{B} \times \left( \mathbf{v} \times \mathbf{B}\right),
\label{eq1}
\end{equation}
where $\mathbf{B}$ is the magnetic field vector and $\mathbf{v}$ is velocity field vector. We are interested in the vertical component of the 
Poynting flux vector $F_y$ (note that, in our notation, the $y$ vector component corresponds to the vertical, while $x$ and $z$ 
correspond to the horizontal directions) which is defined as 
\begin{equation}
F_y=\frac{1}{4\pi} \left( v_y \left( B_x^2 + B_z^2 \right) - B_y \left( v_x B_x + v_z B_z \right) \right) = F_y^\mathrm{v} + F_y^h.
\label{eq2}
\end{equation}
In Eq.~(\ref{eq2}), $F_y^\mathrm{v}$ corresponds to the part of the vertical component of the Poynting flux vector produced by vertical 
motions, and represents the transport of horizontal magnetic field with vertical velocity, while $F_y^h$ is the part of the vertical component, 
generated by horizontal magnetised plasma motions.

\begin{figure*}
\epsscale{1.0}
\plotone{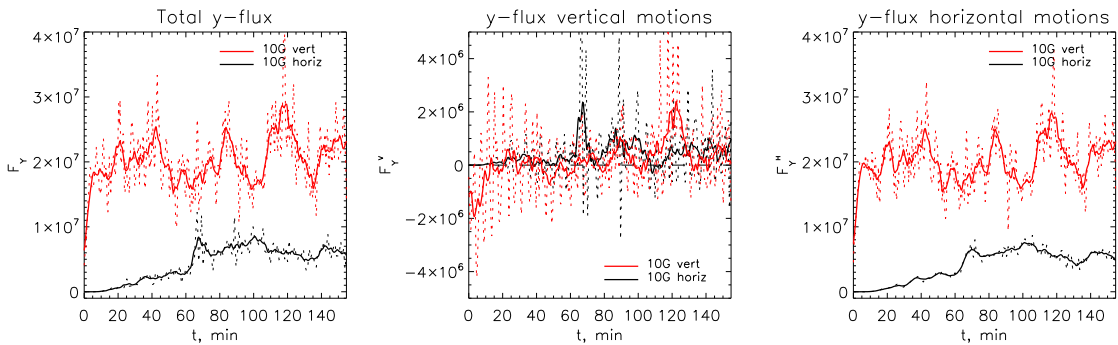}
\caption{The time dependence of the total vertical Poynting flux (left) and its components $F_y^\mathrm{v}$ (middle) and $F_y^h$ (right) 
for the $10~\mathrm{G}$ vertical (red) and $10~\mathrm{G}$ horizontal simulations. Solid lines in the plots correspond to the time 
dependences, obtained by smoothing with a boxcar window of width 5 minutes, while dotted lines are the non-smoothed dependences.
Note that the fluxes for the $10~\mathrm{G}$ horizontal magnetic field simulations are multiplied by a factor of 5 to allow their comparison with
the $10~\mathrm{G}$ vertical simulations.
}
\label{fig4}
\end{figure*}

The $F_y$, $F_y^\mathrm{v}$ and $F_y^h$ components of the Poynting flux vector for $10~\mathrm{G}$ vertical and $10~\mathrm{G}$ horizontal 
magnetic field simulations, measured at $500~\mathrm{km}$ height above the average continuum formation level in the simulation domain, are 
shown in the left, middle and right panels of Fig.~\ref{fig4}, respectively. For both simulations, the total vertical ($y$) component of Poynting flux 
is generally positive. A positive vertical Poynting flux is therefore a general property of the simulated photosphere and does not depend on the 
initial magnetic field geometry. However, our simulations show that the average $F_y$ is about 10 times smaller for the case of horizontal magnetic 
field compared to the $10~\mathrm{G}$ initial vertical magnetic field. This finding holds for the latest stages of the simulation which have achieved 
a state of equilibrium.

\begin{figure*}
\epsscale{1.0}
\plotone{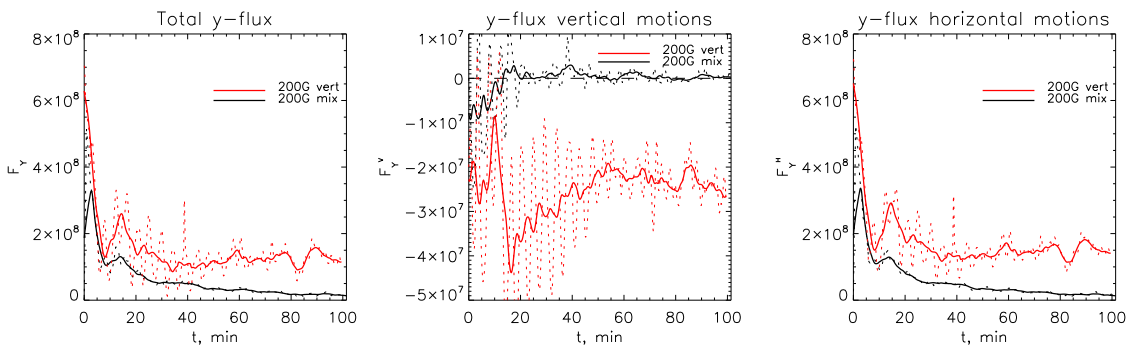}
\caption{Same as Fig.~\ref{fig2} for $200~\mathrm{G}$ initial vertical and $200~\mathrm{G}$ initial mixed polarity magnetic field simulations.}
\label{fig5}
\end{figure*}

The middle panel of Fig.~\ref{fig4} shows the amount of the vertical Poynting flux, generated by the vertical motions. It is evident from the plot that
the horizontal magnetic field simulation produces an average vertical Poynting flux which is positive, while the vertical magnetic field does not.
This confirms the presence of the mechanism of vertical Poynting flux production by advection of horizontal magnetic field in the upward direction, 
as described by \citet{steiner1}. However, as shown in the right panel of Fig.~\ref{fig4}, only about $5\%$ of the total vertical Poynting flux 
is produced by the upward advection of horizontal magnetic fields. This comparison clearly shows that it is the strong vertical magnetic fields in 
the upper photosphere that are mainly responsible for the vertical Poynting flux production. As previously shown by \citet{shelyagangeo}, 
this flux is produced by the interaction of vortex flows in the intergranular lanes with the expanding photospheric magnetic concentrations. The 
positive sign (outward direction) of Poynting flux is not connected either to the initial magnetic field direction, or to the way the magnetic field 
is introduced into the domain. To demonstrate this further, in Fig.~\ref{fig5} we compare the Poynting flux produced with an initial positive 
vertical $200~\mathrm{G}$ field with that of a mixed-polarity vertical $200~\mathrm{G}$ field. The dependence of the total vertical component of 
the Poynting flux with time (left panel of Fig.~\ref{fig5}) shows that after the first 5 minutes, during which the magnetic field is advected into the 
intergranular lanes, the behaviour of the vertical magnetic field simulation stabilises, while the flux emerging from the mixed-polarity simulation 
begins to decay. In contrast to the weaker, $10~\mathrm{G}$ vertical uniform field simulation, the vertical Poynting flux generated by vertical 
motions is now negative, suggesting larger amount of stronger horizontal fields located in the vertical downflow regions. During the first 20 
minutes of the simulation, the mixed-polarity model also shows a negative Poynting flux produced by vertical motions (middle panel of Fig.~\ref{fig5}). 
However, after this stage, when a significant amount of vertical magnetic field has cancelled out, the horizontal small-scale magnetic loop-like 
structures in the granular upflow regions \citep{cameron1} start producing positive Poynting flux, similar to the $10~\mathrm{G}$ horizontal field 
model. As the right panel of Fig.~\ref{fig5} shows, both models generate vertical positive Poynting flux due to horizontal motions in strong 
vertical magnetic field concentrations, which is about two orders of magnitude larger than the Poynting flux produced by vertical motions. 

It has to be noted that the height of $500~\mathrm{km}$ for the measurement of the Poynting flux has been chosen arbitrarily. The results 
presented remain valid everywhere where the Poynting flux is positive, far away from the upper boundary and in a wide range of heights 
starting from about $200~\mathrm{km}$ above the photospheric level. Thus, the influence of the upper boundary condition is considered 
to be small on these measurements.

In general, the $10~\mathrm{G}$ simulations show notably more turbulent behaviour of Poynting flux compared to the $200~\mathrm{G}$ models. 
While the latter shows the oscillations in $F_y$, the former demonstrates a more random dependence of $F_y$  with time. This is due to the 
presence of a large relative amount of weak, turbulent granular fields compared to the strong magnetic field model, where the major part of 
magnetic flux is concentrated in the intergranular lanes.

It is also interesting to note that the Poynting flux in the vertical direction does not scale simply as the square of the magnetic field strength, 
as may be suggested from Eq.~(\ref{eq2}). While the ratio of squares of the mean magnetic field in the $10$ and $200~\mathrm{G}$ 
simulations is 400, the ratio of the average vertical Poynting fluxes in these models is about 10. This is due to the magnetic field in the 
intergranular lanes, which does not increase linearly with increasing initial field. However, it tends to occupy a larger area of the intergranular 
lanes, thus increasing the filling factor. This process continues until the intergranular lanes are fully filled with the magnetic field. Any further 
increases in the initial magnetic field strength lead to the appearance of a dynamically different sunspot umbra region \citep{schumbra}.

\section{Conclusions}
In this paper we carried out a comparative analysis of the Poynting flux generated by simulated magnetised solar photospheres with magnetic 
fields that differ in strength and geometry. We have shown the presence of two different types of vertical Poynting flux generation: one connected 
to the advection of horizontal magnetic field by upflows in the photosphere, while the other is due to horizontal motions of plasma in strong 
intergranular magnetic flux concentrations. Using different initial magnetic field strength photospheric magneto-convection simulations, mixed 
polarity region simulations and simulations with the horizontal magnetic field advected through the bottom domain boundary, we showed 
that the latter mechanism is the preferred one, independent of both how the magnetic field is introduced in the simulated photosphere, and
the direction of initial magnetic field. We demonstrated clearly that the outward flux of electromagnetic energy (Poynting flux) is a general 
property of the simulated photosphere and does not depend on the way how the magnetic field is introduced in the domain.

The initial magnetic field configurations presented in this paper can be related to regions of the solar photosphere. Our weak magnetic field 
($10~\mathrm{G}$) simulations correspond to a case of an extremely weakly magnetised solar granulation, while the $200~\mathrm{G}$ 
uniform and opposite-polarity models allow production of stronger horizontal fields and represent the evolution of unipolar or bipolar regions 
in proximity to sunspots. The results shown suggest that both magnetically active and non-active granulation deliver electromagnetic energy 
to the upper layers of solar atmosphere. While simulated active regions with strong vertical magnetic fields generate 5 times more Poynting 
flux compared to non-active regions, the area covered by non-magnetic granulation can be significantly larger than 
the one covered by active regions, depending on the global solar activity level.

\acknowledgments
\section{Acknowledgement}
This work has been supported by the UK Science and Technology Facilities Council (STFC).

\end{document}